\documentclass[12pt,a4paper]{article}

\usepackage{amsfonts}
\usepackage{amssymb}
\usepackage{graphicx}
\usepackage{amsmath}
\usepackage[dvipdfm]{hyperref}
\usepackage{enumerate}
\usepackage{color}
\RequirePackage{mathrsfs} \RequirePackage[sc]{mathpazo}
\RequirePackage{wasysym} \RequirePackage{setspace}
\begin{document}

\title{\textbf{Critical Behaviors of $3D$ Black Holes with a Scalar Hair}}
\author{A. Belhaj$^{1,2}$, M. Chabab$^2$, H. EL Moumni$^2$, K. Masmar$^2$, M.  B. Sedra$^{3}$ \\
{\hspace{-1cm}\small $^{1}$D\'epartement de Physique, Facult\'e
Polydisciplinaire, Universit\'e Sultan
 Moulay Slimane}\\{\small B\'eni Mellal,  Morocco } \\
{\hspace{-2cm}\small $^{2}$High Energy Physics and Astrophysics
Laboratory, Physics Department, FSSM}\\{\small
 Cadi Ayyad University, Marrakesh, Morocco
} \\
{\small $^{3}$  D\'{e}partement de Physique, LHESIR, Facult\'{e} des
Sciences, Universit\'{e} Ibn Tofail}\\{\small
 K\'{e}nitra, Morocco.} }

\date{ }

 \maketitle


\abstract{
The principal focus of the present work concerns the critical behaviors of a class of three
 dimensional black holes with a scalar field  hair. Since the cosmological constant is viewed as a thermodynamic pressure and its conjugate quantity as a volume, we examine such properties in terms of two parameters  $B$ and  $a$. The latters are
related to the scalar field and the angular momentum respectively.
In particular, we give the equation of state predicting a critical
universal number depending on the $(B,a)$ moduli space. In the
vanishing limit of the $B$ parameter, we recover the usual perfect
gas behavior appearing in the case of the non rotating BTZ black
hole. We point out that in a generic region of the $(B,a)$ moduli
space, the model behaves like a Van der Waals system.}
\\
\\
{\bf Keywords}: 3D black holes, Critical behavior, Thermodynamics.

\newpage
\section{Introduction}
Thermodynamic behaviors of black holes in various dimensions have
received a particular interest through recent important works
[1-18]. The equation of state for certain black holes has been
established using an analysis that involves some characteristics
similar to the Van der Waals $P$-$V$ diagram \cite{0ref,KM,8,Dolan1,Dolan2}.
\\

 The $P$-$V$ criticality of  four dimensional RN-AdS black holes with spherical
configurations have been extensively  studied. In particular,  it
has been shown a remarkable  interplay  between the behaviors of the
RN-AdS black hole systems and the Van der Waals fluids \cite{1ref}.
The $P$-$V$ criticality,  the Gibbs free energy, the first order
phase transition and the behavior near the critical points can be
associated with the statistical liquid-gas systems. On the basis of
\cite{KM}, we have studied the critical behaviors of charged RN-AdS
black holes in arbitrary dimensions of the spacetime\cite{our}. In
fact, we have presented a comparative study in terms of the
dimension and the displacement of the critical points. We have
realized that these parameters can be explored to control the
transition between the small and the large black holes. More
precisely,  we have revealed  that such behaviors vary in terms of
the dimension of the spacetime in which the black hole belongs. A
particular emphasis has been put on the three dimensional case
corresponding to  the BTZ black hole whose critical behaviors are
associated with  the  ideal gas ones. More recently, a novel exact
rotating black hole solution in $(2+1)$-dimensional gravity with a
non-minimally coupled scalar field has been proposed in
\cite{ZXZ,XZ} using an appropriate  metric ansatz.\\
\par
 The study of such   three dimensional blacks can be motivated by
the recent work on the graphene \cite{Meryam}.  The latter is  a two
dimensional
 carbon material  based on   the honeycomb structure having
interesting properties.  It  behaves like a semiconductor with zero
energy gap. Analytically, the corresponding low energy excitations
are described by the Dirac equation dealing with  the 3 dimensional
massless fermions. Roughly, it  has been realized that the curved
graphene sheet with negative constant curvature in the presence of
an external  magnetic source could be modeled using a conformal BTZ
black hole dual solution.\\
\par
The present work concerns the critical behaviors of a class of three
dimensional black holes with a scalar field hair. Since the
cosmological constant is viewed as a thermodynamic pressure and its
conjugate quantity as a volume, we examine such properties in terms
of two parameters  $B$ and  $a$. These parameters correspond to the
scalar field and the angular momentum respectively. Among our
results, we derive the equation of state which predicts a critical
universal number depending on the $(B,a)$ moduli space. If the $B$
parameter is set equal to zero, we recover the usual perfect gas
behavior appearing in the case of the non rotating BTZ black hole.
We point out that in a generic region of the $(B,a)$ moduli space,
the model behaves like a Van der Waals system.\\

\section{The charged non rotating case}
 To start, we consider the following action
\begin{equation}
\mathcal{I}=\frac{1}{2}\int d^3x \sqrt{-g}\left
[R -g^{\mu\nu}\nabla_\mu\phi\nabla_\nu\phi-\xi R\phi^2-2V(\phi)-\frac{1}{4}F_{\mu\nu}F^{\mu\nu}\right]
\end{equation}
where $F = dA$ is the field strength of the $A$ gauge field  and
where $V(\phi)$ is a  scalar potential. In this action,  $\xi$
describes the  coupling  between the gravity and the scalar field.
Following \cite{ZXZ,XZ}  and  choosing $\xi=\frac{1}{8}$, the static
and  the circularly symmetric solution can be obtained from the
following metric
\begin{equation}
ds^2=-f(r) dt^2+ \frac{1}{f(r)} dr^2+r^2 d\psi^2,
\end{equation}
where the space-time variables are $-\infty<t<\infty, r\leq 0$ and
$-\pi\leq\psi\leq\pi$. We also assume that both the scalar field
$\phi$ and the Maxwell field $A_\mu$ depend only on the radial
coordinate $r$. Under such assumptions, the Maxwell equation  gives
\begin{equation}
A_\mu dx^\mu=-Q\ln\left(\frac{r}{r_0}\right)dt
\end{equation}
where $Q$ and $r_0$ are integration constants.  In fact, the  $Q$ is
real and  corresponds
 to the electric charge. While $r_0 > 0$ is associated with  the radial position of the zero
  electric potential surface, which can  be considered as  $+\infty$.

In this context, the metric
reads as
\begin{equation}
f(r)=\left(3\beta-\frac{Q^2}{4}\right)+\left(2\beta-\frac{Q^2}{9}
\right)\frac{B}{r}-Q^2\left(\frac{1}{2}+\frac{B}{3r}\right)ln(r)+\frac{r^2}{\ell^2}
\end{equation}
where $\Lambda$ has been chosen to be $-\frac{1}{\ell^2}$ due to the
fact that in $3D$ the black hole horizon can exist only for a
negative cosmological  constant.  The parameter $B$ corresponds to
the scalar field through $ \phi(r)=\pm\sqrt{\frac{8B}{r+B}}$. It is
worth noting that the parameter $\beta$ is related to  the black
hole mass $M$  and its charge $Q$ as follows
\begin{equation}
\beta=\frac{1}{3}\left(\frac{Q^2}{4}-M\right)
\end{equation}
It is recalled that different models can be obtained by taking
different choices for the function $f$ and the  potential function.
A priori there are many  forms. However,  will consider the same
scalar potential where  it appears the cosmological constant
\cite{ZXZ}
\begin{eqnarray}\nonumber
V(\phi)&=&-\frac{1}{\ell^2}+\frac{1}{512}\left(\frac{1}{\ell^2}+\frac{\beta}{B^2}\right)\phi^6-\frac{1}{18432}\left(\frac{Q^2}{b^2}\right)(192 \phi^2+48\phi^4+5\phi^6)\\
&+& \frac{1}{3}\left(\frac{Q^2}{B^2}\right)\left[\frac{2\phi^2}{(8-\phi^2)^2}-\frac{1}{1024}\phi^6\ln\left(\frac{B(8-\phi^2)}{\phi^2}\right)\right]
\end{eqnarray}
This potential has many nice  cosmological features. We  will just
quote some of them. For very small values of the field, the
potential behaves like $\frac{1}{\ell^2}=\Lambda$  playing  the role
of a (bare) cosmological constant. In principle, the constant
$\Lambda$ can either be positive, zero or negative. However, if we
wish to interpret the solution as a black hole solution, $\Lambda$
will be necessarily negative, because in $(2 + 1)$ dimensions,
smooth black hole horizons can exist only in the presence of a
negative cosmological constant \cite{27}. Moreover, this potential
encodes  the contribution of the self interactions of the scalar
field $\phi$.  For the  uncharged black hole $(Q=0)$, the potential,
up some detail, reduces to  the  $\phi^6$, which matches with black
hole solution in three dimensions.

Now we discuss the thermodynamical quantities associated with the above black hole solution.
 Before  going ahead,
 we should define the Euclidean section $(t\rightarrow
i\tau)$ of the solution and identify the period $\hat{\beta}$ of the
imaginary time with the inverse of temperature \cite{PRD,0}. In
fact, performing the formula for the period,
$\hat{\beta}=\frac{4\pi}{f'(r_+)}$, we get the Hawking temperature.
It is given by
\begin{equation}\label{x2}
T_H=-\frac{\left(B+r_+\right) \left(4 B Q^2 \ell ^2+9 Q^2 r_+ \ell ^2-36 r_+^3\right)}{24
   \pi  r_+^2 \ell ^2 \left(2 B+3 r_+\right)}.
\end{equation}

It follows that  the   corresponding  entropy function  reads as
\begin{equation}
S=\frac{\pi}{2G} r_+.
\end{equation}
To derive the equation of state, we identify the cosmological
constant with the pressure by the relation\cite{state}
\begin{equation}\label{x1}
\ell^2=\frac{1}{8 \pi G P}.
\end{equation}
To get the state equation  $P=P(T,r_+)$, we plug equation
$(\ref{x1})$ in $(\ref{x2})$. Indeed, the direct  calculation
 produces the pressure function as follows
\begin{equation}\label{state}
P=\frac{B Q^2}{72 \pi  r_+^3}+\frac{Q^2}{32 \pi
   r_+^2}+\frac{T}{12 \left(B+r_+\right)}-\frac{T}{6 r_+}.
\end{equation}
This equation is plotted on figure 1.  Comparing the state   equation (\ref{state}) with the one of Van der Waals fluid, we can indetifiy the specific volume
with $v=2\ell_P r_+$, where  $\ell_P$ is the Planck length \cite{KM}

\begin{center}
\begin{figure}[!Ht]
\begin{center}
{\includegraphics[scale=1]{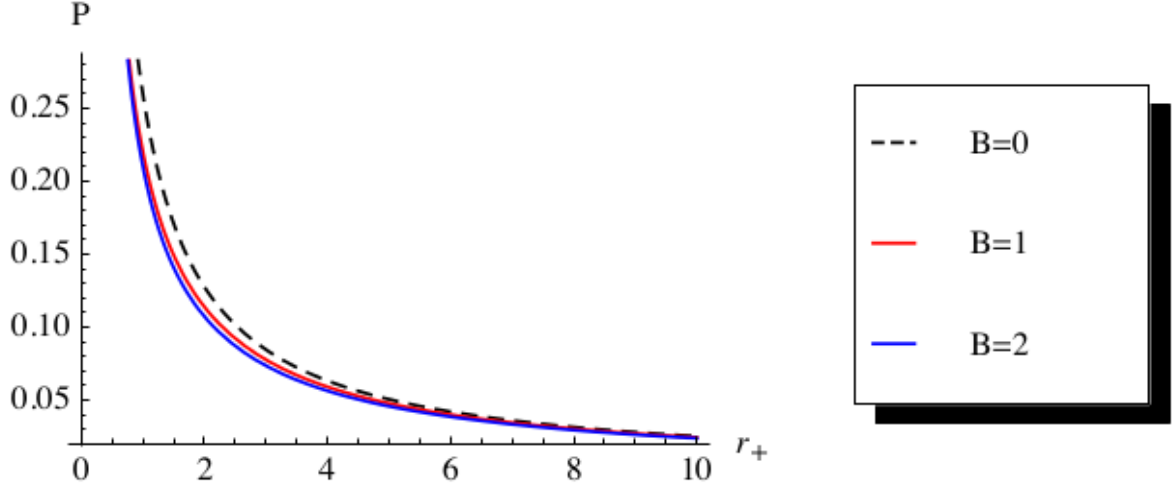}}
\end{center}
 \caption{The $P-r_+$ diagram of charged  $(2+1)$ black hole with the scalar field hair, where  the charge is equal to 1.} \label{fig1} \vspace*{-.2cm}
\end{figure}
\end{center}
It follows from the figure  (\ref{fig1}) that for weak value of horizon radius
the pressure  decreases when the $B$  parameter augments.  Moreover,
for  the scalar field does not bring any physical modification for
large  values  of  $r_+$. This can be understood  from the curves
shown in this figure.

It is also  observed    that the corresponding black hole does not
have any critical behavior. It looks like  an ideal gas. This
situation appears also in the case of BTZ black hole showing a good
agreement with the result obtained in \cite{our}. We expect that
this feature should be true for all no rotating $3D$-black holes.
More general behaviors can be obtained by implementing new physical
parameters.
\section{Uncharged rotating case}
In what follow,   we consider  the $3D$-dimensional gravity with a
non-minimally coupled scalar field.  In the absence of the Maxwell
gauge field,  the model can be described by  the action
\begin{equation}
\mathcal{I_R}=\frac{1}{2}\int d^3x \sqrt{-g}\left[R -g^{\mu\nu}\nabla_\mu\phi\nabla_\nu\phi-\xi R\phi^2-2V(\phi)\right]
\end{equation}
Fixing the  coupling constant $\xi=\frac{1}{8}$, 
 the simplified rotating  black solution can be written  as \cite{ZXZ}
\begin{equation}
ds^2=-f(r) dt^2+\frac{1}{f(r)} dr^2+r^2\left(d\psi^2+\omega(r)dt\right)^2
\end{equation}
where the  functions $f$ and $w$  are given respectively by
\begin{eqnarray}
f(r)&=&3\beta+\frac{2B\beta}{r}+\frac{(3r+2B)^2 a^2}{r^4}+\frac{r^2}{\ell^2}\\
\omega&=&-\frac{(3r+2B)a}{r^3}
\end{eqnarray}
To keep similar analysis given in \cite{XZ}, we consider the same
the potential $V(\phi)$  given by
\begin{equation}
V(\phi)=\frac{1}{512} \left(\frac{a^2 \left(\phi ^6-40 \phi ^4+640 \phi ^2-4608\right)
   \phi ^{10}}{B^4 \left(\phi ^2-8\right)^5}+\phi ^6 \left(\frac{\beta
   }{B^2}+\frac{1}{\ell^2}\right)+\frac{1024}{\ell^2}\right)
\end{equation}

with $\beta=-\frac{M}{3}$.  The  new parameter $a$  is a rotating
parameter  related to the angular momentum. In this case, the
Hawking temperature reads as
\begin{equation}
T_{Ha}=-\frac{3 \left(B+r_+\right) \left(4 a^2 B^2+12 a^2 B r_++9 a^2 r_+^2-8 \pi  P
   r_+^6\right)}{2 \pi  r_+^5 \left(2 B+3 r_+\right)}.
\end{equation}
Based on these quantities,  the equation of state is
\begin{equation}\label{state2}
P=\frac{\left(2 B+3 r_+\right) \left(6 a^2 B^2+15 a^2 B r_++9 a^2 r_+^2+2 \pi
   r_+^5 T\right)}{24 \pi  r_+^6 \left(B+r_+\right)}.
\end{equation}
For  the  small values of the  $B$ parameter,  the equation of state
$(\ref{state2})$ becomes
\begin{equation}\label{xstate}
P=\frac{9 a^2}{8 \pi  r_+^4}+\frac{T}{4 r_+}+B \left(\frac{3 a^2}{2
\pi  r_+^5}-\frac{T}{12 r_+^2}\right)+\mathcal{O}(B^2).
\end{equation}
It is important to note that in the vanishing limit of  the $B$
parameter, the equation of state reduces to an equation describing
the ideal gas. From $(\ref{xstate})$, we get for the vanishing limit
of parameters,
\begin{equation}
P=\frac{T}{4r_+}.
\end{equation}
To transform the equation of state from the  geometric form
$(\ref{xstate})$ to physical one, we use the following redefinition
\begin{equation}
Press=\frac{\hbar c}{\ell_p^2} P,\quad\quad Temp=\frac{\hbar c}{k} T.
\end{equation}
where  the Planck length reads as  $\ell_p^2=\frac{\hbar G_N}{c^2}$.
Multiplying the equation $(\ref{xstate})$ with $\frac{\hbar
c}{\ell_p^2}$,   we find
\begin{equation}
Press=\frac{\hbar c}{\ell_p^2}P=\frac{\hbar
c}{\ell_p^2}\left[\frac{T}{4r_+}+\cdots \right]=\frac{k \;Temp}{4\ell_p^2
r_+}+\cdots.
\end{equation}
If we compare this equation  with the Van der Waals  one
$\left(P+\frac{a}{v^2}\right)(v-b)=kT$,  the  specific volume can be
then identified as\cite{KM}
\begin{equation}
v=4\ell_p^2 r_+.
\end{equation}
It follows also that  the equation of state can be rewritten as
\begin{equation}
P=\frac{T}{v}+\frac{288 a^2}{\pi  v^4}+B \left(\frac{1536 a^2}{\pi
v^5}-\frac{4 T}{3 v^2}\right).
\end{equation}

\begin{center}
\begin{figure}[!ht]
\begin{center}
\includegraphics[scale=1]{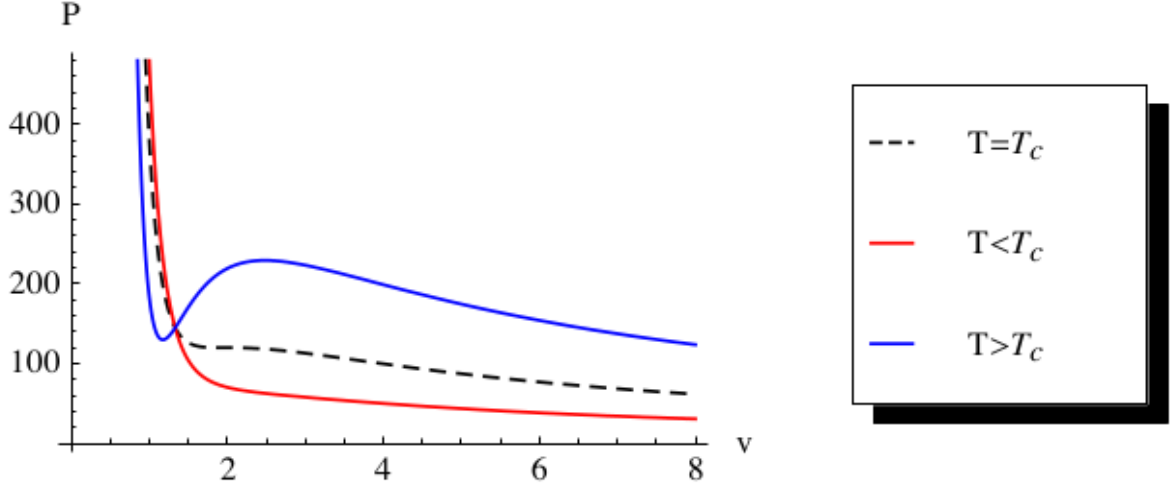}
\end{center}
 \caption{The $P-v$ diagram of rotating $(2+1)$ black holes with an scalar field hair, where $T_c$ is the critical temperature and the charge is equal to 1.} \label{fig2} \vspace*{-.2cm}
\end{figure}
\end{center}

In fact, the discussion on the critical behavior  depends on the
$(B,a)$ moduli space.  At the origin of this moduli space, the
system describes a perfect gas system.  For generic regions of the
moduli space, the corresponding behaviors are shown in figure (\ref{fig2}).

 It  has been observed  that  for $T>T_c$, the
behavior looks like an extended Van der Waals gas.  The  system
exhibits  an inflection point. The corresponding   critical point
ensures a solution of the following conditions
\begin{equation}
\frac{\partial P}{\partial v}=0,\;\;\;\; \frac{\partial^2 P}{\partial v^2}=0.
\end{equation}
The calculation gives the  coordinates of the critical point given
by
\begin{eqnarray}\label{criticcoor}\nonumber
T_c&=&\frac{9 \left(5189+418 \sqrt{154}\right)
   a^2}{50 \pi  B^3},\quad v_c=\frac{4B}{9} \left(\sqrt{154}-8\right),\\
   P_c&=&\frac{59049 a^2 \left(12
   B^4+\left(\sqrt{154}-8\right) B^3-6
   \left(12+\sqrt{154}\right) B+8
   \sqrt{154}+116\right)}{8
   \left(\sqrt{154}-8\right)^5 \pi  B^3}
\end{eqnarray}
From these equations we can see that the existence of the critical
point is controlled by the two parameters $a$ and $B$. The
discussion should be    made in terms of these  two parameters.   In
this way, the $(B,a)$ moduli space together with the displacement of
the  critical point could be used to control the phase transition.
This transition looks like  the transition between  small and the
large black holes.  The corresponding  general study could  be
reported elsewhere.

 A close inspection shows   that the  critical pression  takes negative values in any  generic point
 in the $(a,B)$ moduli space.  Instead of giving an analytic demonstration, we plot  the critical  pression
 in such a moduli space.

\begin{center}
\begin{figure}[!ht]
\begin{center}
\includegraphics[scale=1]{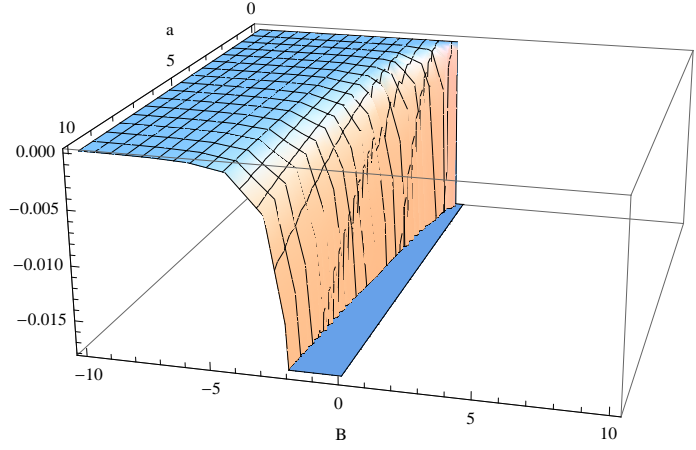}
\end{center}
 \caption{The critical pressure in moduli space $(a,B)$.} \label{fig3} \vspace*{-.2cm}
\end{figure}
\end{center}

It follows from  the figure (\ref{fig3}) that  the critical pressure is always
negative.  In fact, this is not a surprising feature since a similar
thermodynamical situation  is usually  associated with  metastable
states. For
 a generic  temperature, the minimum in the loop  of the van der Waals
 is isotherm at negative values of the pressure. In nature,  such
 a metastable negative pressure might be explored   in  bringing water
to the top of large trees \cite{meta,meta1}.  On the other hand, the situation is analog  to  one associated with the trees.   Inspired from  this example,  we can suggest that the black hole eject the matter from the inside  contrary to black hole physics.   
 In this way, this
negative pressure  behavior could push  the matter outside  of the
horizon. It thus   destroys the black hole object. This can be
supported by the fact that
 a negative pressure  generates  a positive cosmological constant (\ref{x1}).  It is worth noting that
 this natural feature deserves a deep concentration. We will look on it in
 future.

The critical behavior produces the following  universal number
\begin{equation}
\chi=\frac{P_c v_c}{T_c}=\frac{12 B^4+\left(\sqrt{154}-8\right)
   B^3-6 \left(12+\sqrt{154}\right) B+116+8
   \sqrt{154}}{116+8 \sqrt{154}}.
\end{equation}
It is interesting to give some comments concerning this  expression.
First, it  depends  only on  the parameter $B$. More precisely, it
depends on the value of the  scalar field. Second,  for  the
vanishing limit of the $B$ parameter, it can be reduced to the usual
equation $\chi=1$  describing an  ideal gas.

I what follows, it should be interesting to discuss   the stability
of this kind of black hole. In fact, it  has been realized  that the
local stability of a thermodynamic system, with respect to the small
variations of the thermodynamic coordinates, can be investigated  by
the study of the  behavior of the energy, which should be a convex
function of its extensive variable. Moreover,  the positivity of the
heat capacity $C$ is sufficient to ensure the local stability in the
canonical ensemble. In the elaboration of the phase transition, two
essential quantities are the heat capacity ($C$)  showing  thermal
stability for the positive heat capacity and the free energy ($F$)
indicating  the global stability for $F < 0$ \cite{noj1,zwz}. Using
the fact that that the entropy of three dimensional black hole is
related to horizon radius as $S=4\pi r_+$, we can  get
 the heat capacity via the following relation
\begin{equation}
C_a=T\left(\frac{\partial^2 S}{\partial M^2 }\right)=\frac{4 \pi  r_+ \left(B+r_+\right) \left(2 B+3 r_+\right)
\left(r_+^6-a^2 \ell ^2 \left(2 B+3
   r_+\right){}^2\right)}{a^2 \ell ^2 \left(2 B+3 r_+\right){}^3 \left(4 B+3 r_+\right)+r_+^6 \left(4 B^2+3
   r_+ \left(2 B+r_+\right)\right)},
\end{equation}
with $M$ is the mass of the black hole. In the case od non rotating,
 this quantity  is reduced to
\begin{equation}
C=\frac{4 \pi  r_+ \left(B+r_+\right) \left(2 B+3 r_+\right)}{4 B^2+3 r_+ \left(2 B+r_+\right)}
\end{equation}
To understand such a behavior, we plot the corresponding  quantities
in term of $r_+$ for different values of $a$.
\begin{center}
\begin{figure}[!ht]
\begin{center}
\includegraphics[scale=1.5]{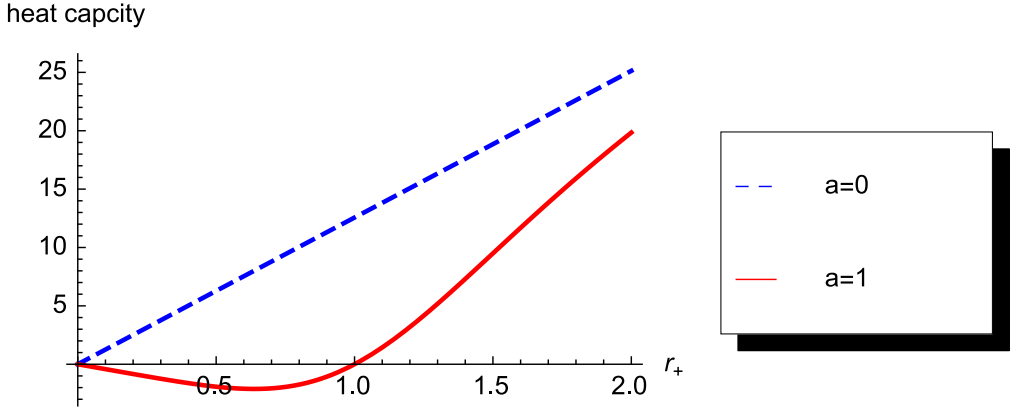}
\end{center}
 \caption{The heat capacity for $B=1$ and $\ell=1/3$.} \label{fig3} \vspace*{-.2cm}
\end{figure}
\end{center}
It follows that,  for  the non rotating  black hole solution, the
heat capacity is always positive showing the  thermal stable.
However,  the introduction of the rotation parameter destabilizes
the black hole. This can be seen from  from the figure $4$ where the
heat capacity is negative in the region $r_+ \leq 1$. This study, in
fact, deserves deeper study to  investigate the  thermodynamical and
statistical properties of such a black hole. We hope to come back to
this issue  in  future works

\section{Conclusion}
To conclude,  we have investigated the  thermodynamical behaviors of
a class of $3D$-black holes with a scalar hair, in the presence of
the  cosmological constant considered as a thermodynamic pressure
and its conjugate quantity as a volume.  We have shown that the
corresponding equation of state predicts a critical universal number
depending on the $(B,a)$ moduli space. In the absence of the $B$
parameter,  the usual perfect gas behavior appearing in the case of
non rotating BTZ black hole has been  obtained. We have remarkably
pointed out a  good agreement of our analysis with the results given
in \cite{ZXZ}.

It is worth noting that this study comes up with many open
questions. In fact, the calculation has been made only  for some
thermodynamical quantities. To give a   complete picture, others
quantities should be implemented, including free energy, partition
function an so on.

 In connection with the graphene,  many efforts have been
devoted to the study of the opening energy gap  using different
approaches including the application of the implementation of   the
magnetic field.  Since in three dimensions, the gauge field is dual
to a scalar, the black holes presented in this paper  could open
windows for  such graphene activities.  This is an important issue
that deserves a special investigation. In connection with these
activities,  \cite{evap} has  shown that quantum-corrected BTZ BH
may evaporate or else anti-evaporate similarly to 4D Nariai BH as is
observed by Bousso and Hawking.  It is should be interesting to
investigate the effect the introduction of the scalar hair on this
context.


\newpage
\begin{thebibliography}{99}
\bibitem{0ref} B. P. Dolan, {\em
Pressure and volume in the first law of black hole thermodynamics},
Class. Quantum Grav. {\bf 28} (2011) 235017.

\bibitem{1ref} A. Chamblin, R. Emparan C. V. Johnson3, R. C.
Myers, {\em  Charged AdS black holes and catastrophic holography},
Phys. Rev. {\bf D 60}(1999) 064018.

\bibitem{KM} D. Kubiznak and  R. B. Mann, {\em P-V criticality of charged AdS black holes},  JHEP {\bf 1207} (2012) 033.
\bibitem{GKM} S. Gunasekaran, D. Kubiznak, R. B. Mann, {\em  Extended phase
space thermodynamics for charged and rotating black holes and
Born-Infeld vacuum polarization}, {\tt arXiv:hep-th/1208.6251v2}.
\bibitem{DKKMT} B. P. Dolan, D. Kastor, D. Kubiznak, R. B. Mann, J. Traschen, {\em
Thermodynamic Volumes and Isoperimetric Inequalities for de Sitter
Black Holes},  {\tt arXiv:1301.5926}.

\bibitem{30} S. Hawking and D. N. Page, {\em Thermodynamics of Black Holes
in Anti-de Sitter Space},   Commun.Math.Phys. {\bf 83} (1987) 577.
\bibitem{kenitra} A. Belhaj, K. Bilal, A. El Boukili, M. Nach, M.B. Sedra, {\em Solutions and
thermodynamics of non commutative Liouville black hole},
Int.J.Geom.Meth.Mod.Phys.(2013).


\bibitem{chin1} C. Song-Bai, L. Xiao-Fang, L. Chang-Qing, {\em $P$-$V$ Criticality of an AdS Black Hole in f(R) Gravity},
 Chin. Phys.Lett Vol.{\bf 30}, No.6(2013)060401.

\bibitem{our} A. Belhaj, M. Chabab, H. El Moumni, M. B. Sedra, {\em
On Thermodynamics of AdS Black Holes in Arbitrary Dimensions}, Chin. Phys.Lett
Vol.{\bf 29}, No.10(2012)100401.
\bibitem{4} A. Chamblin, R. Emparan, C. Johnson and R. Myers, {\em  Charged
AdS black holes and catastrophic holography},  Phys.Rev. {\bf D60}
(1999) 064018.

\bibitem{central}  C. Wang,Y. Gui,  {\em Rediscussion of charged dilaton-axion black hole entropy},
Cent. Eur. J. Phys.
, Volume {\bf 7}, (2009) Issue 3.


\bibitem{5} A. Chamblin, R. Emparan, C. Johnson and R. Myers,
{\em  Holography, thermodynamics, and fluctuations of charged AdS
black holes}, Phys.Rev. {\bf D60}  (1999) 104026.

\bibitem{50} M. Cvetic, G. W. Gibbons, D. Kubiznak and C. N. Pope, {\em Black Hole Enthalpy and an
Entropy Inequality for the Thermodynamic Volume}, Phys. Rev. {\bf D
84}, 024037 (2011), {\tt arXiv:1012.2888 [hep-th]}.

\bibitem{6} B. P. Dolan, D. Kastor, D. Kubiznak, R. B. Mann and J. Traschen, {\em Thermodynamic
Volumes and Isoperimetric Inequalities for de Sitter Black Holes},
{\tt  arXiv:1301.5926 [hep- th]}.

\bibitem{7} B. P. Dolan, {\em Pressure and volume in the first law of black hole
thermodynamics}, Class. Quant. Grav. {\bf 28}, 235017 (2011), {\tt
[arXiv:1106.6260 [gr-qc]}.

\bibitem{8}J, Liang and B. Liu,  {\em Thermodynamics of noncommutative geometry inspired BTZ black hole based on Lorentzian smeared mass
distribution}, EPL. {\bf 100} (2012) 30001.

\bibitem{epl1} H. L. Li, S. Z. Yang, {\em Hawking radiation from the charged BTZ black hole
with backreaction}, EPL.{\bf 79}(2007)20001.

\bibitem{epl2} S. Chakraborty, N. Mazumder, R. Biswas, {\em The generalized second law of thermodynamics and the nature
of the entropy function}, EPL. {\bf 91} (2010) 40007.

\bibitem{Dolan1}  D. O'Connor, B. P. Dolan, M. Vachovski, {\em Critical Behaviour of the Fuzzy Sphere},{\tt arXiv:1308.6512
}




\bibitem{Dolan2} B.P. Dolan,
{\em The compressibility of rotating black holes in D-dimensions}
{\tt arXiv:1308.5403 }





\bibitem{ZXZ} L. Zhao, W. Xu, B. Zhu, {\em  Novel rotating hairy black hole in (2+1)-dimensions}, Commun.Theor.Phys. {\bf 61} (2014) 475-481, {\tt
arXiv:1305.6001}.
\bibitem{XZ}
 W. Xu, L. Zhao, {\em  Charged black hole with a scalar hair in (2+1)
dimensions},Phys.Rev. {\bf D87} (2013) 12, 124008, {\tt arXiv:1305.5446}.
\bibitem{Meryam}
M. Cvetic, G. W. Gibbons. {\em Graphene and the Zermelo Optical
Metric of the BTZ Black Hole M}, {\tt  arXiv:1202.2938}.
\bibitem{PRD}
G. W. Gibbons,  S. W. Hawking, {\em  Action integrals and partition
functions in quantum gravity}, Phys. Rev. {\bf D 15}, (1977) 2752.

\bibitem{0} D. Grumiller, W. Kummer, D.V. Vassilevich, {\em Dilaton Gravity in Two
Dimensions},  Phys.Rept.{\bf 369}(2002)327, {\tt
arXiv:hep-th/0204253}.
\bibitem{state} B.D. Brian, {\em
The cosmological constant and the black hole equation of state},
Class. Quantum Grav. 28 (2011) 125020, {\tt arXiv:1008.5023}.

\bibitem{chil} D.Kastor, S. Ray, J. Traschen, {\em The Enthalpy and the Mechanics of AdS Black Holes},  Class.Quant.Grav.
 {\bf26} (2009) 195011, {\tt arXiv:hep-th/0904.2765}.


\bibitem{27} D. Ida, {\em No black hole theorem in three-dimensional gravity}, Phys. Rev. Lett. {\bf 85} (2000) 3758,
{\tt arXiv:gr-qc/0005129}.

\bibitem{meta} R. G. Mortimer, {\em Physical
Chemistry,
Third Edition},{\tt Elsevier Academic Press.}

\bibitem{meta1} C. Domb, J. Lebowitz, {\em Phase transition and critical phenomena,
 Vol-15}, {\tt Academic Press.}

\bibitem{meta2} William E. Brower, Jr. and David J. Schedgick, L.
 Kimball Bigelow, {\em Thick Glassy Water by Liquid Quenching on a Diamond Wafer},
  J. Phys. Chem B {\bf 106} 4565 (2002)

\bibitem{meta3} T. Loerting et al, PCCP {\bf 3} 5355 (2002)




\bibitem{noj1}  Shin'ichi Nojiri, Sergei D. Odintsov, {\em Quantum (in)stability of 2D charged dilaton black holes and 3D rotating black holes}, Phys.Rev.{\bf D59}:044003,1999, {\tt arXiv:hep-th/9806055}

\bibitem{zwz} A. Sheykhi, S. H. Hendi,  S. Salarpour, {\em Thermodynamic stability of BTZ dilaton black holes}, {\em Phys. Scr. {\bf 89} (2014) 105003}.

\bibitem{evap} Shin'ichi Nojiri, Sergei D. Odintsov, {\em Can Quantum-Corrected BTZ Black Hole Anti-Evaporate?}, Mod.Phys.Lett.{\bf A13}:2695-2704,1998. {\tt arXiv:gr-qc/9806034}.

\end{thebibliography}
\end{document}